\documentclass{jfm}
\usepackage{graphicx}
\begin{document}
\pubyear{2002} \volume{453} \pagerange{427-438}
\title{Coalescence of sessile drops}
\author[C. Andrieu, D. A. Beysens, V. S. Nikolayev and Y.Pomeau]
         {C.\ns A\ls N\ls D\ls R\ls I\ls E\ls U$^1$,\ns D.\ns A.\ns B\ls E\ls Y\ls S\ls E\ls N\ls S$^1$,\ns
         V.\ns S.\ns N\ls I\ls K\ls O\ls L\ls A\ls Y\ls E\ls V$^1$\ns \\[\affilskip]
         \and Y.\ns P\ls O\ls M\ls E\ls A\ls U$^2$}
\affiliation{$^1$ESEME, Service des Basses Temp\'eratures, CEA Grenoble, France\thanks{Mailing address:
CEA-ESEME, Institut de Chimie de la Mati\`ere Condens\'{e}e de Bordeaux, 87, Avenue du Dr. Schweitzer, 33608
Pessac Cedex, France; email: dbeysens@cea.fr}\\[\affilskip]
$^2$Laboratoire ASCI, Bat. 506, 91405 Orsay Cedex, France}
\date{\today}
\maketitle

\begin{abstract}
We present an experimental and theoretical attempt to describe the kinetics of coalescence of two water drops
on a plane solid surface. The case of the partial wetting is considered. The drops are in an atmosphere of
nitrogen saturated with water where they grow by condensation and eventually touch each other and coalesce. A
new convex composite drop is rapidly formed that then exponentially and slowly relaxes to an equilibrium
hemispherical cap. The characteristic relaxation time is proportional to the drop radius $R^\ast $ at final
equilibrium. This relaxation time appears to be nearly 10$^{7}$ times larger than the bulk capillary
relaxation time $t_{b}=R^{\ast }\eta /\sigma $, where $\sigma $ is the gas-liquid surface tension and $\eta $
is the liquid shear viscosity.

In order to explain this ``giant'' relaxation time, we consider a model that involves an Arrhenius kinetic
factor resulting from a liquid-vapor phase change in the vicinity of the contact line. The model results in a
large relaxation time of the order $t_{b}\exp(L/{\cal R}T)$ where $L$ is the molar latent heat of
vaporization, $\cal R$ is the gas constant and $T$ is the temperature. We model the late time relaxation for
a near spherical cap and finds an exponential relaxation whose typical time reasonably agrees with the
experiment.
\end{abstract}

\section{Introduction}

Fusion or coalescence between drops is a key process in a wide range of phenomena: phase transition in fluids
and liquid mixtures or polymers, stability of foams and emulsions, or sintering in metallurgy (Eggers 1998).
It is why the problem of coalescence has already received considerable attention. Most of the studies of this
process have been devoted so far to the coalescence of two spherical drops floating in a medium. The kinetics
of the evolution before and after the drops have touched each other is governed by the hydrodynamics inside
and outside the drops and by the van der Waals forces when the drops are within mesoscopic distance from each
other (Yiantsios \& Davis 1991). As a matter of fact, the composite drop that results from the coalescence of
two drops relaxes to a spherical shape within a time which is dominated by the relaxation of the flow inside
and outside the drops (Nikolayev \textit{et al.} 1996, Nikolayev \& Beysens 1997). There are no studies, to
our knowledge, of the coalescence of two sessile drops after they touch each other. In this paper, we report a
preliminary study of the dynamics and morphology of this process, in the case of hemispherical water droplets
which grow slowly on a plane surface at the expense of the surrounding atmosphere. This forms what is called
``dew'' or ``breath figures'' (Beysens \textit{et al.} 1991, Beysens 1995).
The drops eventually touch each
other and coalesce to form an elongated composite drop, which then relaxes to a spherical cap or to a less
elongated drop. We study the composite drop relaxation, which depends on the droplet size and morphology
(contact angle $\theta $). When the contact angle is large ($70^{\circ }<\theta <90^{\circ }$), the composite
drop relaxes very rapidly (within a time frame of a video camera) to a hemispherical cap (Zhao \& Beysens
1995). The relaxation process also looks fast if the contact angle is small ($\theta <20^{\circ}$). However,
in this case the drop never relaxes to a hemispherical cap. Its shape remains always complicated. We study
here the intermediate range of the contact angles where the composite drop relaxes to a near-hemispherical
drop. We report on two experiments where the contact angles are $30^\circ$ and $53^\circ$. We find that the
relaxation process can be described by an exponential function, with a typical time proportional to the
radius of the drop.

Such a relaxation process cannot be accounted by the bulk dissipation of the flow inside the drops. Available
information in the literature shows that there are no widely accepted theories because of the complicated
interplay between macroscopic and molecular scale phenomena (Blake \& Haynes 1969, de Gennes 1985, Dussan \&
Davis 1986, Hocking 1994). We propose here a simple approach based on an Arrhenius factor coming from the
gas-liquid transition that occurs in the vicinity of the contact line and that limits the contact line
mobility.

\section{Experimental}

When water vapor condenses on a cold substrate under partial-wetting conditions, one observes an assembly of
droplets (dew) that continuously grow and merge. We use this process to study the coalescence of sessile
drops. The surface properties of the substrate play a crucial role in this process through the contact angle.
The experiments consists in observing with a microscope the drops that grow at a low rate in a condensation
chamber. The images before and after drop coalescence are studied by means of an image analysis system. The
condensation chamber involves a Peltier-element thermostat to lower the substrate temperature. The chamber is
a plexiglas box filled with $\mathrm{{N}_{2}}$ gas saturated with water at room temperature ($22^{\circ }$C).
To avoid dust and to saturate the gas, $\mathrm{{N}_{2}}$ is bubbled in pure water. The gas flow is
controlled with a flowmeter. All the experiments presented here are performed with $\mathrm{{ N}_{2}}$
saturated with water at the flow rate of 13.5~$\mathrm{{cm^{3}/s}}$. The substrate on which condensation
takes place is set on a 5~mm thick block made of electrolytic copper. This block ensures a homogeneous
temperature diffusion from the Peltier element to the substrate. The thermal contact is ensured by a silicon
free heat sink compound (Tech Spray). The temperature is monitored with a thermocouple of type K placed on
the copper block near the substrate. The experimental procedure consists in rapidly cooling the substrate
from room temperature down to $15^{\circ }$C. The growth of dew is observed with an optical microscope and
recorded with a CCD camera. The video data are analyzed by a digital processing system.

The surface of condensation is made of 0.7~mm thick glass wafers with 37~mm diameter. The surfaces are first
cleaned with diluted fluorhydric acid, then with optical soap and finally rinsed with pure water and ethanol.
The cleaned substrates are baked up at $120^{\circ }$C, dipped in a
fluorochlorosilane (FClSi) solution and then put back in an oven at $%
120^{\circ }$C for an hour. At this temperature the FClSi molecules are
chemically grafted on the glass surface and do not react any more with
water. Several wafers were used for the experiments reported here.

The treated surface is hydrophobic. Water condenses on it as droplets with a finite contact angle $\theta $.
To measure $\theta $, a small drop of water (2~mm diameter) is deposited with a syringe on the substrate and
observed with a CCD camera through a macrolens. We report two sets of experiments for $\theta =30^\circ$ and
$53^{\circ }$. (The contact angle was made different by changing the experimental conditions.) The hysteresis
of the contact angle was of order $15^{\circ }$ in both cases.

\begin{figure}
  \begin{center}
  \includegraphics[height=6cm]{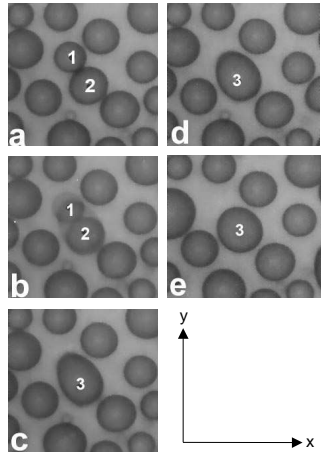}
  \end{center}
\caption{Photos of the coalescence process. The side of each photo corresponds to 276 $\mu$m. (a): t=5.59~s;
(b): t= 5.63~s, the coalescence time; (c): 5.65~s; (d): 6.65~s; (e): 42.76~s.} \label{1}
\end{figure}

The morphology of the coalescence-induced composite drop is analyzed as
follows. After calibration, we measure the contact area, the major and minor
axes (the second one is taken perpendicularly to the first) for several
``composite'' drops as a function of time. In Fig.~\ref{1}
we report a typical evolution of two hemispherical drops (``parents'') that coalesce to form an elongated,
composite drop (``child''), which eventually returns to a spherical cap shape (within 3--4~\%). The return to
a spherical cap indicates that the surface roughness is small enough so as the contact line is not pinned in
a metastable state. Note the Fig.~\ref{1}b where, within a time-frame (40~ms), one can observe the two
parents and the child at the same time. This is because the video image is made of two interlaced frames with
20~ms scanning time.

\section{Observation and preliminary analysis}

In such a process, we can analyze the behavior of several quantities of interest.

\subsection{Position of the center of mass}

\begin{figure}
  \begin{center}
  \includegraphics[height=6cm]{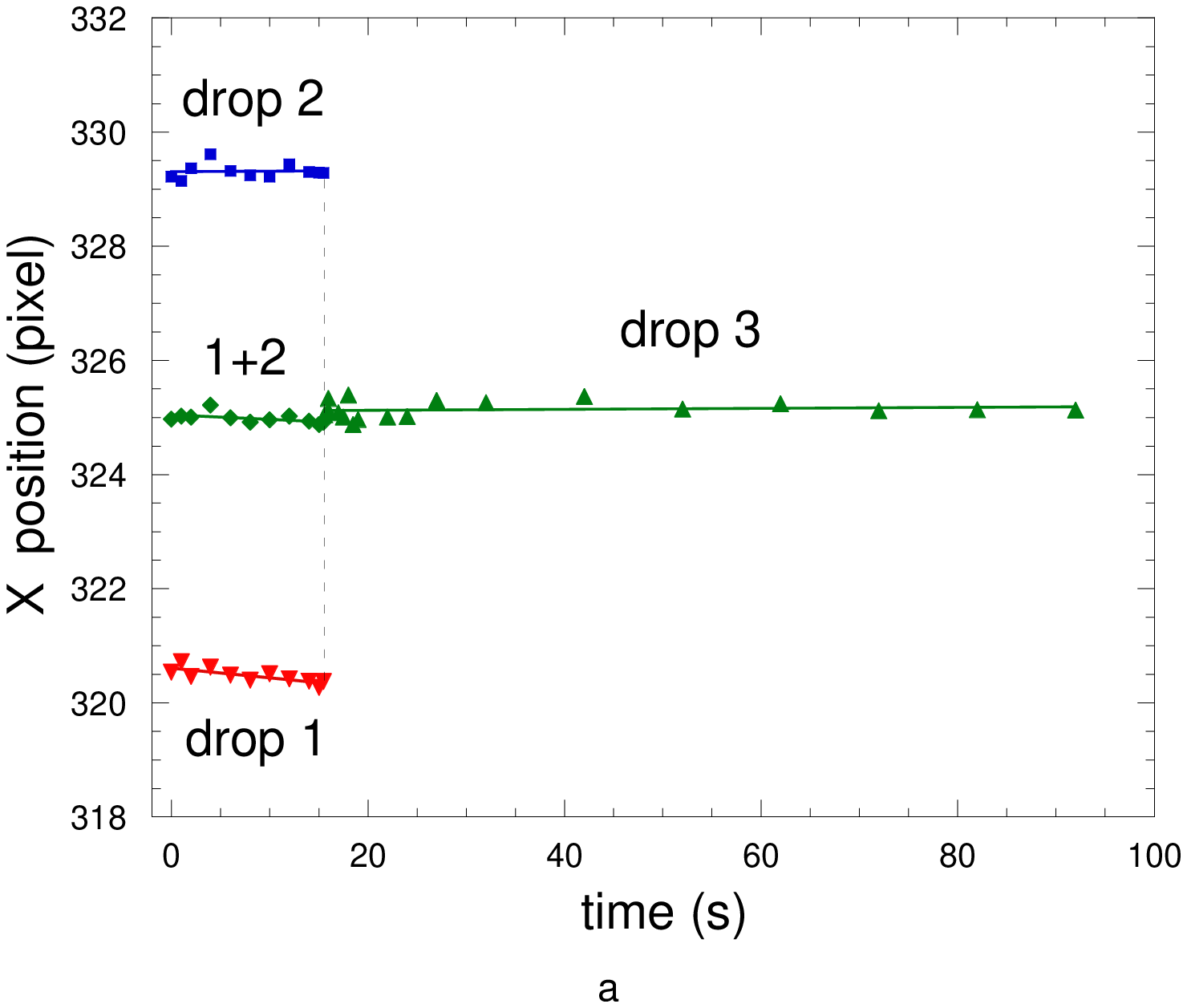}\hspace*{7mm}\includegraphics[height=6cm]{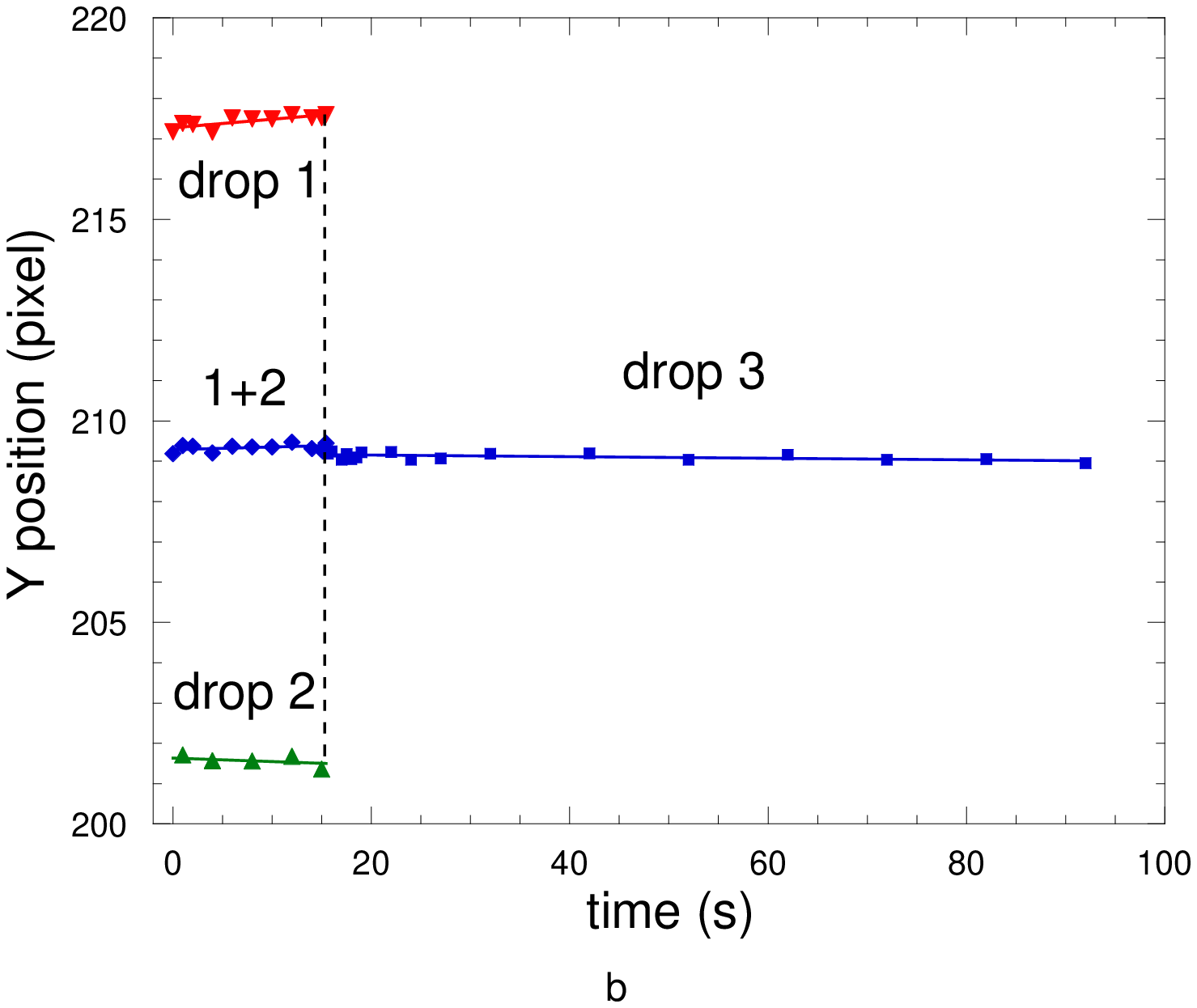}
  \end{center}
\caption{Evolution of the coordinates of the centers of mass of the parent drops (1, 2) and of the composite
drop 3 after coalescence. (a) abscissas (b) ordinates} \label{2}
\end{figure}

We report in Figs.~\ref{2}
the details of a typical coalescence event, similar to the one shown in Fig.~\ref{1}. Here the parent drops
are labeled 1 and 2. Drop 3 is the child drop. The coordinates ($x_{i}$, $y_{i}$) of the center of mass of
the drops $i = 1, 2$ and 3 are calculated from their contact area on the substrate. For drops 1 and 2, and
the late times of drop 3, drops are nearly hemispherical and this approximation is fully justified. A
difficulty comes when dealing with the early times of drop 3 that shows only one symmetry axis and whose 3-D
shape cannot be determined from the pictures. However, its asymmetry with respect to the axis perpendicular
to the symmetry axis is small. Therefore, we assume that the position of the center of mass can be calculated
from the 2D formulas as if the drop were of homogeneous thickness. The calculation of the center of mass of
the system drop 1 + drop 2 can be performed according to the well-known formulas (here the contact angle does
not enter in the equation, because it is assumed to be the same for both drops):
\begin{eqnarray}
x_{3} &=&\frac{x_{1}R_{1}^{3}+x_{2}R_{2}^{3}}{R_{1}^{3}+R_{2}^{3}},
\label{mass} \\
y_{3} &=&\frac{y_{1}R_{1}^{3}+y_{2}R_{2}^{3}}{R_{1}^{3}+R_{2}^{3}},
\end{eqnarray}
where $x_{i}$, $y_{i}$, $R_{i}$ are the coordinates and radius of drop $i$,
with $i=1,2,3$.

Before coalescence, during growth, the coordinates of the center of mass of the parents are reported in
Figs.\ref{2} with respect to the axes shown in Fig.\ref{1}. We also report the evolution of the center of
mass of the ensemble drop 1 + drop 2, calculated from the above Eq.\ref{mass}. A small variation is observed.
We attribute it to the vicinity of neighboring drops, which distorts the gradient of water concentration
around them. (Note that the mass transfer is maximum at the triple line, where the temperature gradient is
maximum (Steyer \textit{et al.} 1992).

As already noted, the center of mass of drop 3 does not move during its relaxation towards a spherical cap.
We note, however, a systematic change of position between the coordinates of the drop 3 center of mass and
the center calculated from the ensemble drop 1 + drop 2. This deviation can be related to external forces,
such as those that pin the contact line. As a matter of fact, the center of mass after coalescence is shifted
to the largest drop (drop 2), that is, the drop showing the longest contact line.

\subsection{Relaxation of the composite drop}

\begin{figure}
  \begin{center}
  \includegraphics[height=4cm]{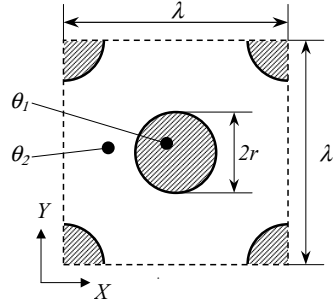}
  \end{center}
\caption{ Sketch of the coalescence event as in Figs.~\ref{1}. (a), (b), (c) denote the time sequence.}
\label{3}
\end{figure}

\begin{figure}
  \begin{center}
  \includegraphics[height=6cm]{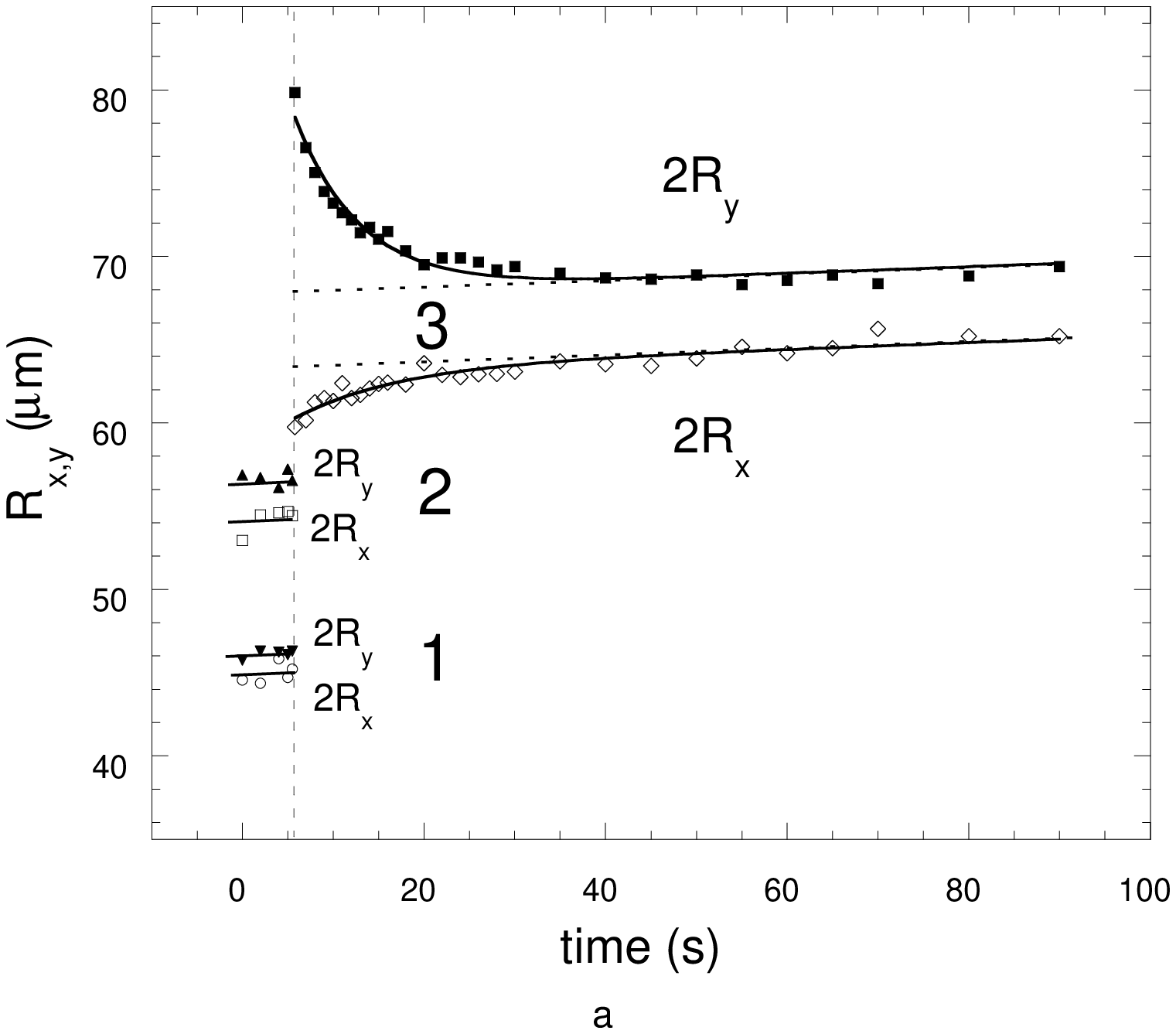}\hspace*{7mm}\includegraphics[height=6cm]{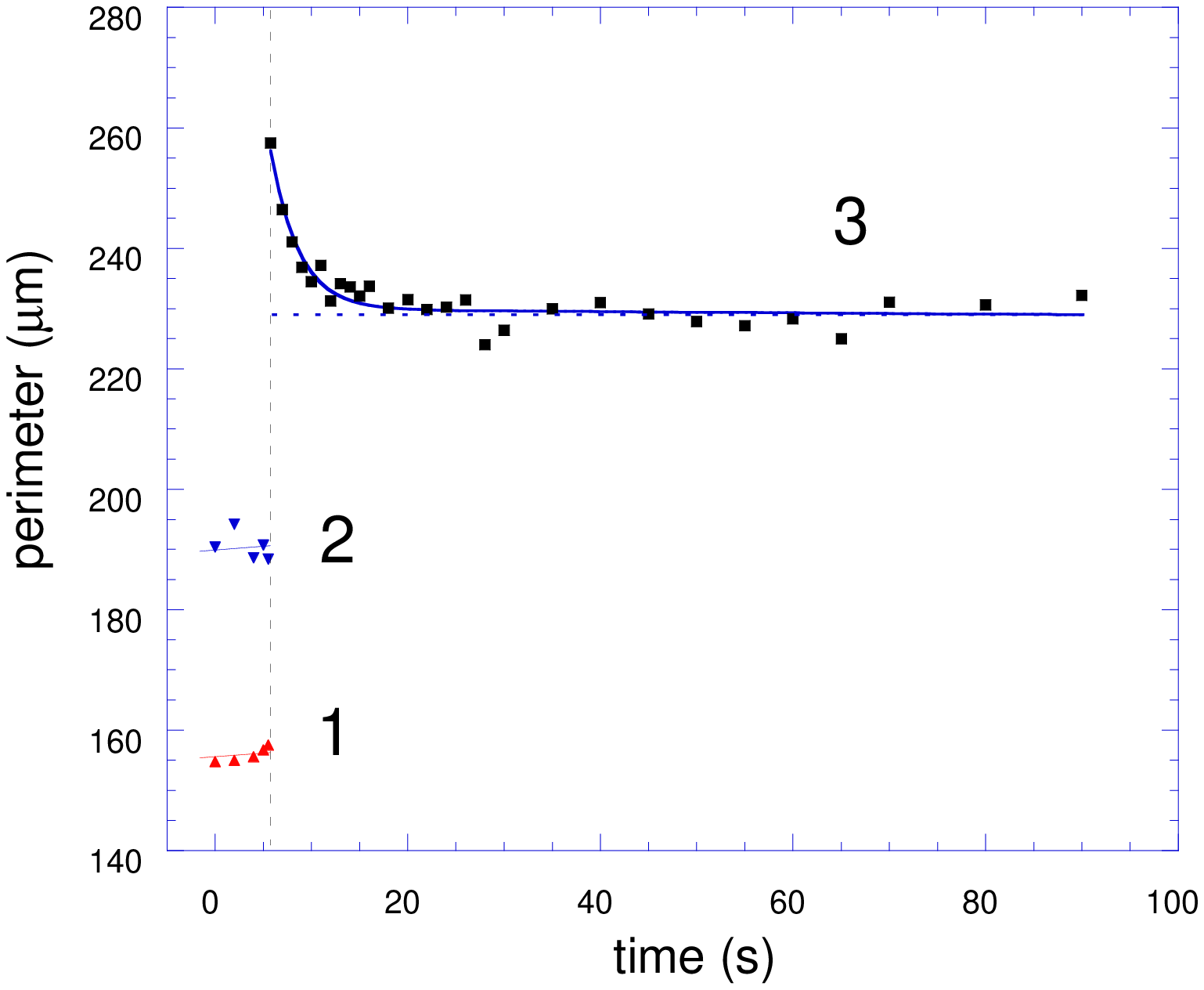}\\
  \includegraphics[height=6cm]{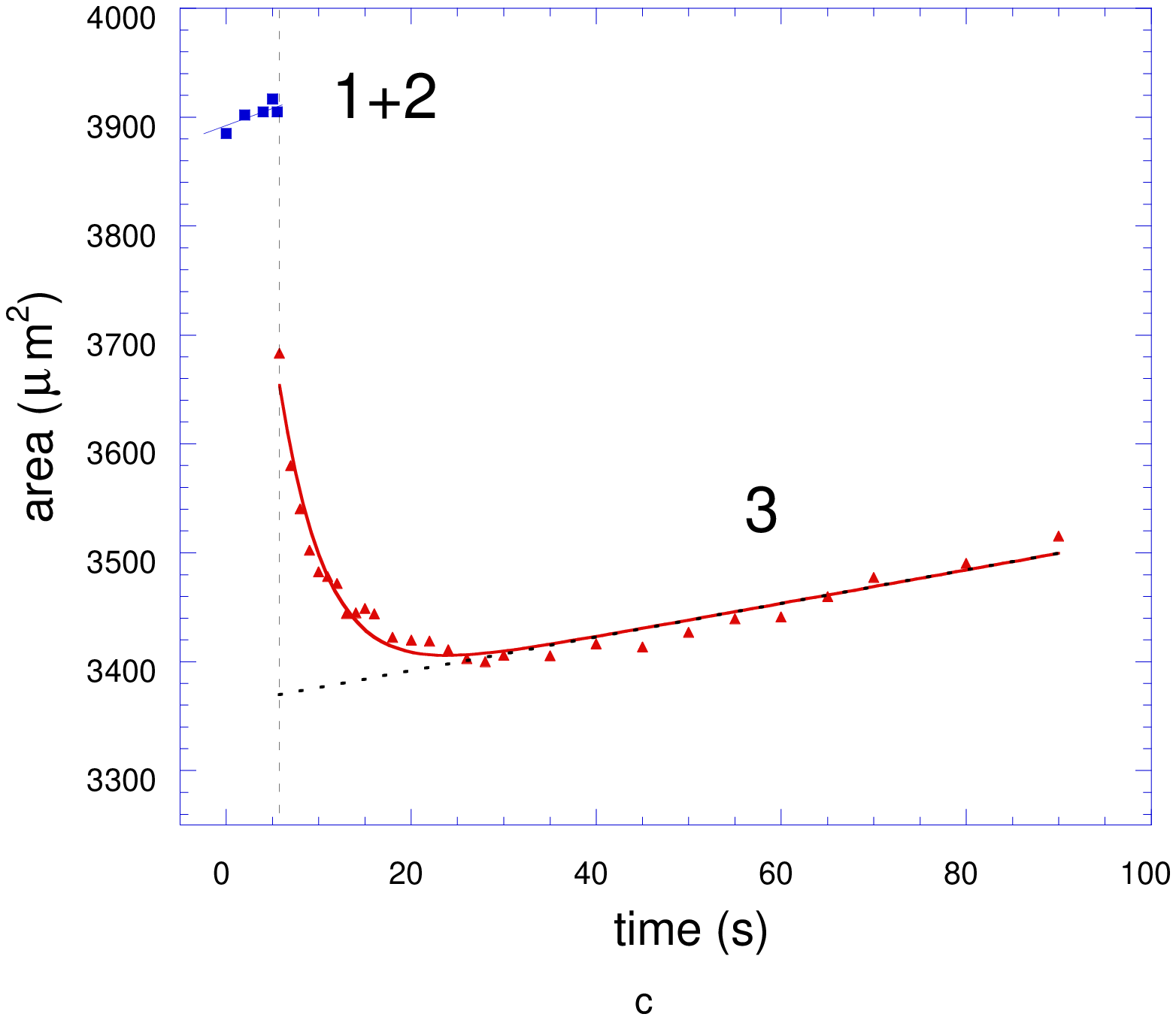}\hspace*{7mm}\includegraphics[height=6cm]{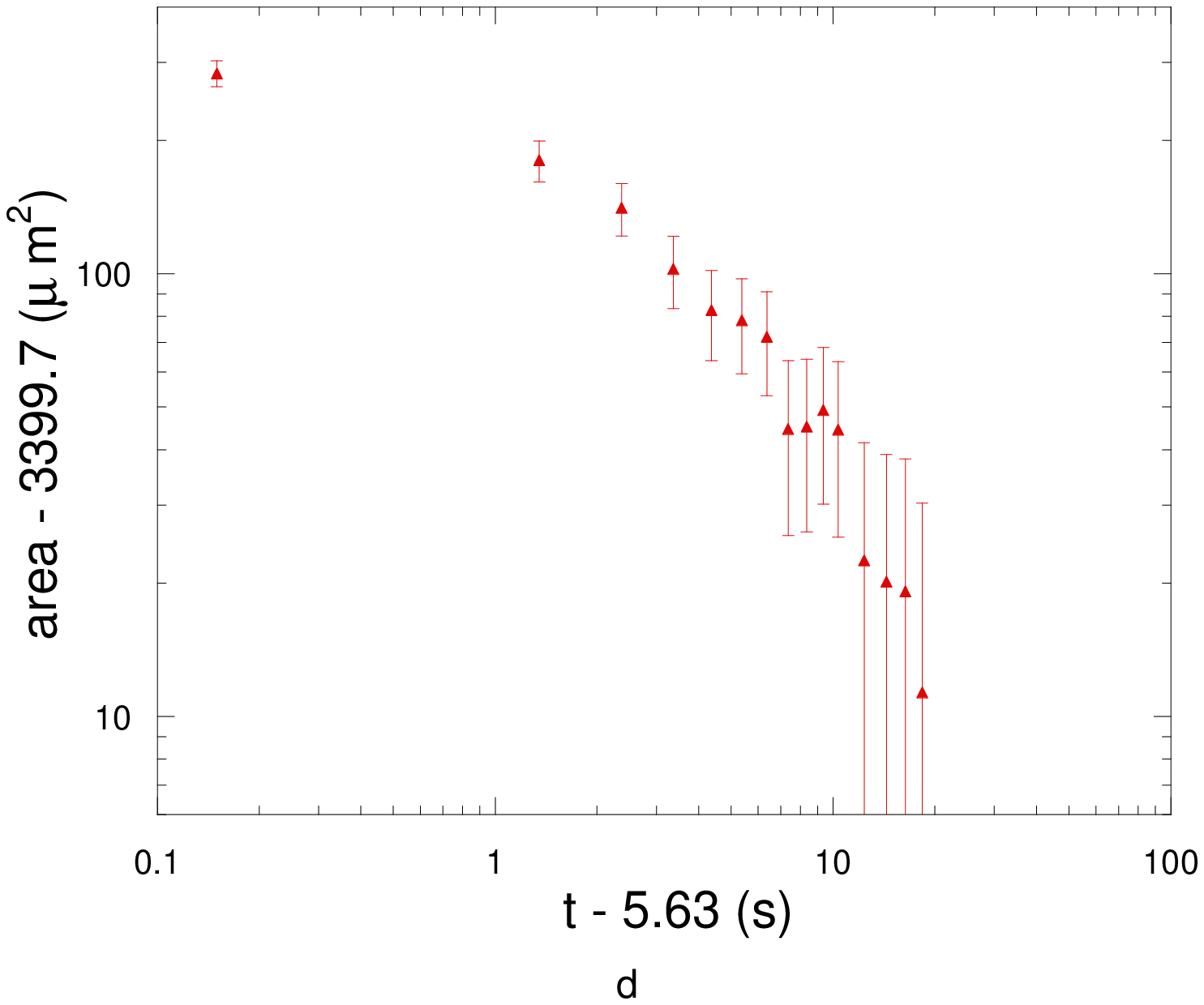}
  \end{center}
\caption{ Evolution of the (a) large and small axes, (b) perimeter, and (c) area of the  drops 1 and 2
("parents", see Figs.~\ref{1}), and drop 3 (composite drop). The curves are fits to Eq.~\ref{Fit}. The
relaxation times for these three fits are equal within the experimental error. The (dotted) asymptotic lines
correspond to the condensation-induced slow growth. In (d) the data from (c) is shown in a log-log plot.}
\label{4}
\end{figure}

Fig.~\ref{3} shows a schematic view of the coalescence process.
In a simplified manner, the contact line of the child, or composite, drop can be assumed to be an ellipse of
width $R_{x}$ (resp. $R_{y}$) along the small (resp. long) axes. The most obvious dynamical quantity to
analyze is the evolution of $R_{x}$ and $R_{y}$. A typical curve is shown in Fig.~\ref{4}a,
showing as well that the parent drops are
circular within $\pm 3\%$ accuracy. The coalescence process is characterized
by three time regions:
\begin{enumerate}
\renewcommand{\theenumi}{\roman{enumi}}
\item Nucleation of a liquid bridge between the two parent drops and subsequent
formation of a convex composite drop occurs in a time period smaller than the half video scanning time, i.~e.
20 ms (Fig.~\ref{1}b). In the theoretical explanation sketched below, the corresponding dynamics is fast
because, during the growth of this bridge, the contact line moves locally by vapor condensation, a process
not slowed down by an \textit{a priori} Arrhenius factor.\label{(i)}

\item Decrease of the large axis $R_{y}$ with time, small increase of the
small axis $R_{x}$, such as the ratio $R_{y}/R_{x}$ eventually reaches a value about unity.\label{(ii)}

\item Slow growth of the drop due to condensation.\label{(iii)}
\end{enumerate}
The second regime (\ref{(ii)}) takes most of the time of the coalescence process. Once the evolution of
$R_{x}$ and $R_{y}$ is corrected from the slow growth observed in (\ref{(iii)}), the evolution of the small
axis $R_{x}$ is most often negligible and the evolution of the drop can be characterized by $R_{y}$ only.

Parameters other than $R_{y}$, as the liquid-solid contact area $A_{LS}$
(Fig.~\ref{4}c) or its perimeter $P$ (Fig.~\ref{4}b), can give a useful
information. Their behavior is quite similar to that of $R_{y}$. In the
following, we will discuss only the evolution of the contact area $A_{LS}$, a
quantity which can be measured with the best accuracy. As shown in
Fig.~\ref{4}c, the evolution of $A_{LS}(t)$ cannot be described by a power law.
It can, however, be successfully fitted to the following function
\begin{equation}
A_{LS}(t)=A_{0}\exp [-(t-t_{0})/t_{c}]+A_{1}(t-t_{0})+A_{2}.  \label{Fit}
\end{equation}
The first term corresponds to the relaxation of the composite drop, which dominates in regime (\ref{(ii)})
and the second and third term to the growth by condensation detected during period (\ref{(iii)})
(Fig.\ref{4}d). Since the stage (\ref{(i)}) is very rapid, the time $t_{0}$ can be considered as the time of
the beginning of the coalescence. The free parameters in the fit are $t_{c}$, $A_{0}$, $A_{1}$, and $A_{2}$.
The area $A_{2}$ is related to the drop radius $R^{\ast}$ at equilibrium (when the drop becomes a spherical
cap) by $A_{2}=\pi R^{\ast^{2}}$.

There are three time scales that correspond to the regimes (\ref{(i)}), (\ref{(ii)}) and (\ref{(iii)}). The
first regime is controlled by the time scale $< 20$~ms. The second regime corresponds to $t_c\sim 5$~s.
Before the beginning of the third stage, the relaxation to the final shape is almost finished. On this third
stage, the total liquid mass grow due to condensation, which is necessary to make the drops grow and
coalesce. This third time scale is of the order of 150~s and can be characterized by the coefficient $A_1$.
Since these three time scales differ strongly, the corresponding regimes can be considered independently from
each other. In the following, we discuss only the time $t_{c}$ that fully characterizes the relaxation of the
convex drop. Its value should be independent of the externally imposed small condensation rate.

\begin{figure}
  \begin{center}
  \includegraphics[height=6cm]{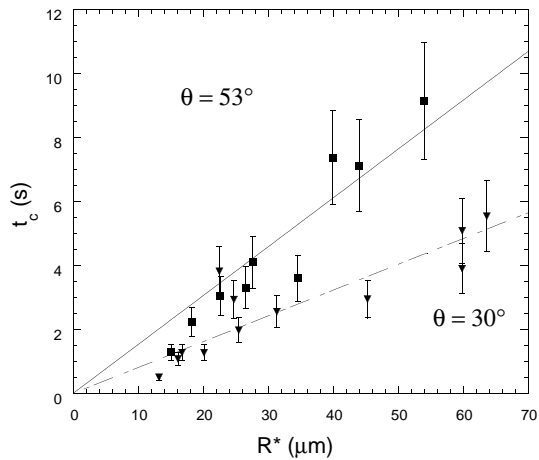}
  \end{center}
\caption{ Variation of the relaxation time $t_c$ versus the composite drop radius $R^\ast$ measured at the end
of relaxation, see Fig. \ref{3}. Data corresponding to $\theta=30^\circ$ (triangles) and $\theta=53^\circ$
(squares) are shown. The lines are the best fits.} \label{5}
\end{figure}

In Fig.~\ref{5}, we report how $t_{c}$ vary with $R^{\ast }$ for two contact angles ($\theta =53^{\circ }$
and $30^{\circ }$).
The data show that $t_{c}$ varies linearly with $R^{\ast }$ and fits to
\begin{equation}
t_{c}=\frac{1}{u}R^{\ast },
\end{equation}
where $u$ is homogeneous to a velocity. This fit gives $u=(6.5\pm 0.4)\,10^{-6}$~m/s for $\theta =53^{\circ
}$ and $u=(12\pm 1)\,10^{-6}$~m/s for $\theta =30^{\circ }$. The statistical error corresponds to the values
of the linear correlation coefficient 0.93 ($\theta =53^{\circ}$) and 0.87 ($\theta =30^{\circ }$).

Note that $t_{c}$ should depend on the difference between the sizes of the two parent drops. However, this
dependence is weak when the difference is small because the composite drop is nearly symmetrical, see
Fig.~\ref{1}.

The relaxation time appears to be larger for $\theta =53^{\circ}$ than for $\theta =30^{\circ }$. This looks
paradoxical since the relaxation driving force (see Eq.~\ref{effeteta} below) should be larger for larger
angles thus causing a smaller relaxation time. (As a matter of fact, for large angles $\theta>70^\circ$ the
relaxation becomes too fast to be measured within a video scanning time, which compares well with this
argument.) The small relaxation time at small angles can be explained by the pinning of the contact line on
surface defects, leading to a metastable equilibrium in a shorter time, so that the total relaxation time
becomes smaller. The pinning is likely to occur for a smaller contact angle where the relaxation driving
force is smaller. For $\theta\le 20^\circ$ the shape of the drops becomes quite complex and thus unsuitable
for analysis because of the contact line pinning (Zhao \& Beysens 1995).

The detailed study of the contact angle dependence of $t_c$ is beyond the scope of this paper, which is
focused on the unexpected long relaxation time found in these experiments.

\section{Theoretical analysis}

Let us consider first the relaxation of the drop to the spherical cap equilibrium shape as described by bulk
hydrodynamics. The rate of relaxation of the slightly deformed drop is defined by three characteristic times.
The first of them is the inviscid inertial time $t_{e} = (\rho\,R^{\ast \,3}/\sigma )^{1/2}$, where $\rho$ is
the liquid density. The time $t_{e}$ accounts for the slowing down of the relaxation by the inertia of the
liquid. The second is the viscous inertial time $t_{i}=R^{\ast \,2}/\nu$, where $\nu $ is the kinematic
viscosity, which accounts for dissipation at high Reynolds number in the boundary layer. The third
characteristic time is the time $t_b$ of the viscous relaxation driven by the surface tension $\sigma $. It
is given by the expression $t_b=R^{\ast }\eta /\sigma $, where $\eta $ is the shear viscosity.

Considering the data for water at $20^{\circ }$C ($\eta =10^{-3}$~Pa~s, $\nu =10^{-6}$~m$^{2}$/s, $\sigma
=73$~mN/m), it is easy to check that all of these times calculated for water are many orders of magnitude
less than the relaxation time reported in Fig.~\ref{5} (nearly seven orders for the viscous flow
approximation). This section deals with the explanation of this huge difference.

The relationship between the contact line motion and phase change was studied previously in a number of
works. The reaction of the liquid meniscus on the externally imposed contact angle change was analyzed by
Wayner (1993). He assumed implicitly that the vapor condenses homogeneously over the whole meniscus thus
causing the contact line to advance along the already prewetted solid surface. We note that the subsequent
studies (Anderson \& Davis 1994, Nikolayev \& Beysens 1999, Nikolayev \textit{et al.} 2001) showed that the
phase change rate varies along the meniscus being much larger in the contact line vicinity than on the other
part of the meniscus. This effect was taken into account by Anderson \& Davis (1995) who studied the
influence of the evaporation on the value of the dynamic contact angle. The contact line was assumed to move
due to the non-zero liquid-solid slip coefficient.

We note that the previous studies of the volatile liquids relied on the already known mechanisms of the
contact line motion (de Gennes 1985) proposed for the non-volatile liquids. These mechanisms cannot explain
the very small factor $10^{-7}$ observed in our experiments.

The contact angle hysteresis that appears due to the surface defects is small and results in a small deviation
of the final drop shape from that of the spherical cap, which would form in the absence of the hysteresis.
This deviation can be estimated from Fig. \ref{4}a as $(R_y-R_x)/(R_y+R_x)\approx 3\%$, $R_x$ and $R_y$ being
taken for $t\gg t_c$. The analysis of Nikolayev \& Beysens (2001) shows that the week defects are not likely
to cause such a strong change in the relaxation rate. This huge difference calls for a radical change in
theory.

It is well known that such very small non-dimensional factors appear very often in the presence of an
activation process that gives very small Arrhenius non-dimensional factors. For the contact line motion, it
has been proposed recently that such a factor should be present (Pomeau 2000) because of the following
remark: the contact line, sticking on the solid, cannot move by hydrodynamic motion, since the fluid velocity
is zero on the solid (no-slip condition). This condition can be satisfied during the contact line motion when
the evaporation (when the liquid is receding) or condensation (when the liquid is advancing) occurs in the
vicinity of the contact line (Seppecher 1996). Although the rate of this phase change can be large (it is
proportional to the contact line speed), it does not result in a strong mass change of the drop because of
the small part of the area where the phase change occurs. The rate of this process is therefore independent of
the small global condensation rate present in our experiments.

Hereafter, we shall consider the receding process only, that is, evaporation. This evaporation is a thermally
activated process because molecules in the liquid are in the bottom of a potential well, due to the
attraction of the other molecules, an attraction necessary to keep the cohesion of the liquid against
spontaneous self-evaporation. Therefore, the rate of evaporation should be proportional to a very small
Arrhenius factor
\begin{equation}
K=\exp(-\frac{W}{k_{B}T}),  \label{KEW}
\end{equation}
where $W$ is the difference of potential energy between the liquid and vapor
side, taken as positive (practically, this potential energy is zero in the
dilute vapor, and $-W$ in the liquid). Supposing that the potential energy
grows monotoneously from a well in the liquid to its zero value in the vapor,
one would get for $W$ the latent heat per molecule. Using now the molar latent
heat $L\approx 44$ kJ/mol (which is its value for $20^\circ$C) one obtains
$K=\exp(-L/{\cal R}T)\sim 10^{-8}$, where $\cal R$ is the ideal gas constant.

This physical phenomenon bears some similarity with the Molecular Kinetics Theory (Blake \& Haynes 1969),
that also involves the Arrhenius factor. Note that, in contrast to the Molecular Kinetics theory, the
Arrhenius factor $K$ defined above is independent of the difference between the actual contact angle and its
equilibrium value (and so is intrinsically very small and independent of the actual contact angle, as
observed).

Now, it becomes simple to formulate the equations of motion for the relaxation of shape of the droplet: the
pure hydrodynamic phenomena (inertial or viscous) are so fast that they can be taken as having reached
equilibrium for a given droplet contour on the solid. In other terms, all the dynamics is ``enslaved'' to the
slowest process, the receding motion of the contact line driven by evaporation. Therefore the dynamics is as
follows: given a contour $\Gamma$ for the droplet on the solid (a closed curve, i.~e. a triple contact line),
and the volume of the droplet (its change of volume either by evaporation near the contact line or by
condensation from the supersaturated vapor is a relatively small effect on the time scale of the shape
relaxation), one computes the shape of the droplet by solving (numerically in general) Laplace's equation for
the surface of the droplet $z=z(x,y)$:
\begin{equation}
\frac{\partial ^2z}{\partial x^2}\left[1+\left(\frac{\partial z}{\partial y}\right)^2\right]+
\frac{\partial^2z} {\partial y^2} \left[1+\left(\frac{\partial z}{\partial x} \right)^2\right]-2{
\frac{\partial ^2z}{\partial x\partial y}\frac{\partial z}{ \partial x}\frac{
\partial z}{\partial y}}=p\;\left[1+\left(\frac{\partial z}{\partial x} \right)^2+\left(\frac{
\partial z}{\partial y}\right)^2\right]^{3/2}  \label{lap}
\end{equation}
where $p$ is the Lagrange multiplier. It allows the volume of the droplet
consistent with the Dirichlet boundary condition $z=0$ on $\Gamma $ to be
imposed. This way of solution corresponds to the assumption that the process
of the relaxation of the drop surface controlled by the very small
hydrodynamic time scale $t_e$ is much faster that the contact line motion
controlled by the (large) time scale $t_b/K$. The solution of this (well
posed) mathematical problem for $z(x,y)$ does not satisfy in general the
Young-Dupr\'{e} condition $\theta =\theta _{eq}$ for this contact angle. It
gives \textit{a priori} a non constant value, $\theta(s)$, along $\Gamma $,
a function of the curvilinear abscissa such that
\begin{equation}
\cos \theta (s)=\left.\left[1+\left(\frac{\partial z}{\partial x}\right)^2+\left(\frac{
\partial z}{\partial y}\right)^2\right]^{-1/2}\right|_{z=0}
\end{equation}

The next step consists in using a mobility relation for the slow dynamics of the contact line, that in
general takes the form
\begin{equation}
v_{n}=K\,\frac{\sigma }{\eta }\,F(\theta ,\theta _{eq})  \label{mobility}
\end{equation}
where $\sigma /\eta$ is the molecular velocity scale, of the order of the velocity obtained from the
viscosity/capillarity balance, and $K$ is the (small) Arrhenius factor introduced before. Moreover, $F$ is a
non-dimensional function of the contact angle, of order $1$ and equal to zero at equilibrium, when $\theta
=\theta _{eq}$. In this framework, the driving force for the contact line motion is the difference between the
actual value of the contact angle and its equilibrium value. The velocity $v_{n}$ is the local speed of the
contour, perpendicular to it along the solid plane, driven by evaporation, not by fluid motion. This kind of
relationship is not new of course, and expressions like
\begin{equation}
F\propto(\cos \theta _{eq}-\cos \theta )  \label{effeteta}
\end{equation}
have been written before.

It is a major endeavor, far beyond the scope of the present work, to study in details the specific problem of
droplet merging in this theoretical (and well defined) framework. In particular, a complication arises
because the droplet contour is receding somewhere and advancing elsewhere. The advancing motion involves
condensation at the drop foot instead of evaporation during receding. Since there is no potential barrier for
the condensing molecules (unless there is a film barrier on the surface of the liquid), one expects advancing
to happen at molecular speed, with $K\approx 1$. However, if there are receding parts on $\Gamma$, the
slowest receding process is still dominating the whole relaxation. Indeed, one expects that when two droplets
merge, the major effect will be a receding motion of the contour, that is much larger at the beginning than
at the end, when equilibrium is reached.

In what follows, we shall sketch the analysis of the most simple situation, and assume that the droplet is
circular and recedes from a large spherical cap to its equilibrium shape with a smaller contact circle.
Indeed, there is no change of shape of the droplet in this very simple case, contrary to what happens in the
experiments. (This question of the shape change is discussed in detail by Nikolayev \& Beysens 2001.) For the
circular droplets, the solution of Laplace's equation is simple, and yields the following value (constant
along $\Gamma$, a circle now) for $\theta $:
\begin{equation}
\theta =\arcsin \left( \frac{r}{R}\right) ,
\end{equation}
where we consider contact angles smaller than $\pi /2$, as in the experiments. The radius $r$ is the radius
of the contour $\Gamma $, and $R$ is the radius of curvature of the droplet surface (a spherical cap in the
present case). The radii $r$ and $R$ are related by the condition for the drop volume:
\begin{equation}
V=\frac{\pi }{3}\left( R-\sqrt{R^{2}-r^{2}}\right) \left[ r+R^{2}-R\sqrt{R^{2}-r^{2}}\right]
\end{equation}
Eq. \ref{mobility} yields an ``explicit'' equation of motion for $r(t)$:
\begin{equation}
\frac{dr}{dt}=K\frac{\sigma }{\eta }F(\theta (r),\theta _{eq})
\end{equation}
This equation is explicit because $\theta (r)$ can in principle be obtained
from the volume condition. We shall make another simplifying assumption,
namely that the contact angle is small, so that $R$ is much larger than $r$,
an assumption that results in
\begin{equation}
V\cong\frac{\pi r^{4}}{4R}
\end{equation}
The contact angle becomes $\theta \cong r/R\cong 4V/(\pi r^{3})$. The equation of motion for $r$ then reads
\begin{equation}
\frac{dr}{dt}=K\frac{\sigma }{\eta }F\left( \theta (r)=\frac{4V}{\pi r^{3}},\theta _{eq}\right).
\end{equation}
Assuming now that the function $F$ is linear in $\theta -\theta _{eq}$ when this difference is small, one
gets:
\begin{equation}
\frac{dr}{dt}=K\frac{\sigma }{\eta }\left( \frac{4V}{\pi r^{3}}-\theta_{eq}\right).   \label{fineq}
\end{equation}
Note that there is a positive sign in front of the right hand side in the equation above: when the contact
line is receding, $\theta $ is less than its equilibrium value, making negative $[4V/(\pi
r^{3})-\theta_{eq}]$, so that $r$ decreases. This equation can be integrated explicitly, yielding a
cumbersome expression. To outline the calculation, let us introduce the non-dimensional quantity $\omega
=r/R^\ast$, which becomes equal to $1$ when the contact angle reaches its equilibrium value. The equilibrium
drop radius is $R^{\ast }=(4V/\pi \theta _{eq})^{1/3}$. The time dependence of $\omega $ follows from the
implicit relation:
\begin{equation}
\int \frac{d\omega \omega ^{3}}{1-\omega ^{3}}=-\frac{w}{\pi R^{\ast }}t, \label{fineq2}
\end{equation}
where $w=K\sigma /\eta $ is the typical speed of the contact line that includes the Arrhenius factor $K$. The
integration over $\omega $ can be performed. We shall detail only the asymptotic approach to equilibrium,
corresponding to $\omega $ getting close to $1$ from above:
\begin{equation}
\omega \approx 1+C\exp (-\frac{3wt}{\pi R^{\ast }}),
\end{equation}
where $C$ is a constant that depends on the initial condition. It is interesting to notice that this
simple-minded calculation allows the exponential decay observed in the experiments to be recovered. With
$K\sim 10^{-8}$, the relaxation time
\begin{equation}
t_{c}=\frac{\pi }{3}\frac{R^{\ast }}{w}=(\frac{\pi }{3}\frac{\eta }{K\,\sigma })R^{\ast }  \label{TC}
\end{equation}
is almost of the same order as in the experiments.

\section{Concluding remarks}

The surprising finding that the relaxation of coalescing drops on a substrate can occur with a typical
timescale that is many order of magnitude larger than bulk hydrodynamics is a spectacular demonstration of
the strong dissipation that occurs in the motion of the triple (solid-liquid-gas) contact line. We have good
reasons to deem that this dissipation is due to a gas-liquid phase transition at the portion of the
vapor-liquid interface adjacent to the triple line. This dissipation introduces a very small $K\sim 10^{-8}$
Arrhenius factor in the relation between the driving force and the contact line velocity. According to our
idea, the dissipation appears when the contact line recedes so that local evaporation takes place. Further
theoretical and experimental work seems necessary to fully describe the evolution of such a coalescence
phenomenon.

\begin{acknowledgments}
We thank the CEA/DAM in Limeil for their help with the surface coating. \end{acknowledgments}


\begin{thebibliography}{}
\bibitem[1]{b1}\textsc{Anderson, D. M. \& Davis S. H.} 1994 Local heat flow near contact lines.
\emph{J. Fluid. Mech.} \textbf{268}, 231 -- 265.
\bibitem[2]{b2}\textsc{Anderson, D. M. \& Davis S. H.} 1995 The spreading of volatile liquid droplets on heated
surfaces. \emph{Phys. Fluids} \textbf{7}, 248 -- 264.
\bibitem[3]{b3}\textsc{Beysens, D., Steyer, A., Guenoun, P., Fritter, D. \&
Knobler, C.\ M.\ }1991 How does dew form, \emph{Phase Transitions} \textbf{31}, 219 -- 246.
\bibitem[4]{b4}\textsc{Beysens, D.} 1995 The formation of dew. \emph{Atmospheric Research}
\textbf{39}, 215 -- 237.
\bibitem[5]{b5}\textsc{Blake, T. D. \& Haynes, J. M.} 1969 Kinetics of liquid/liquid Displacement. \emph{J. Colloid
Interface Sci.} \textbf{30}, 421 -- 423.
\bibitem[6]{b6}\textsc{van Dussan, E. B. \& Davis, S. H.} 1986 Stability in systems with moving contact lines.
\emph{J. Fluid Mech.} \textbf{173}, 115 -- 130.
\bibitem[7]{b7}\textsc{Eggers, J.} 1998 Coalescence of Spheres by Surface Diffusion. \emph{Phys. Rev. Lett.}
\textbf{80}, 2634 -- 2637.
\bibitem[8]{b8}\textsc{de Gennes, P.-G.} 1985 Wetting: statics and dynamics. \emph{Rev. Mod. Phys.} \textbf{57}, 827 -- 863.
\bibitem[9]{b9}\textsc{Hocking, L. M.} 1994 The spreading of drops with intermolecular forces, \emph{Phys. Fluids}
\textbf{6}, 3224 -- 3228.
\bibitem[10]{b10}\textsc{Nikolayev, V. S., Beysens, D. \& Guenoun, P.} 1996 New hydrodynamic mechanism for drop coarsening.
\emph{Phys. Rev. Lett.} {\bf 76}, 3144  -- 3148.
\bibitem[11]{b11}\textsc{Nikolayev, V. S. \& Beysens, D.} 1997 Hydrodynamically limited
coalescence. \emph{Phys. Fluids}  \textbf{9}, 3227 -- 3234.
\bibitem[12]{b12}\textsc{Nikolayev, V. S. \& Beysens, D. A.} 1999 Boiling crisis and non-equilibrium drying transition.
{\it Europhysics Letters} {\bf 47}, 345 -- 351.
\bibitem[13]{b13}\textsc{Nikolayev, V. S. \& Beysens, D. A.} 2002 Relaxation of nonspherical sessile drops towards equilibrium.
{\it Phys. Rev. E}, {\bf 65}, 046135.
\bibitem[14]{b14}\textsc{Nikolayev, V. S., Beysens, D. A., Lagier, G.-L. \& Hegseth, J.} 2001 Growth of a dry spot under
a vapor bubble at high heat flux and high pressure. {\it Int. J. Heat Mass Transfer}, \textbf{44}, 3499 --
3511.
\bibitem[15]{b15}\textsc{Pomeau, Y.} 2000 Representation of the moving contact line in the equations of fluid
mechanics, \emph{Comptes Rendus Acad. Sci., Serie IIb} \textbf{238}, 411 -- 416.
\bibitem[16]{b16}\textsc{Seppecher, P.} 1996 Moving contact lines in the Cahn-Hilliard theory, \emph{Int. J. Eng. Sci.}
\textbf{34}, 977 -- 992.
\bibitem[17]{b17}\textsc{Steyer, A., Guenoun, P. \& Beysens, D.} 1992 Spontaneous jump of droplets, \emph{Phys. Rev. Lett.}
\textbf{68}, 64 -- 66.
\bibitem[18]{b18}\textsc{Wayner, P. C.} 1993 Spreading of a liquid film with a finite contact angle by the
evaporation/condensation process. \emph{Langmuir} \textbf{9}, 294 -- 299.
\bibitem[19]{b19}\textsc{Yiantsios, S. G. \& Davis, R. H.} 1991 Close Approach and Deformation of Two Viscous Drops due
to Gravity and van der Waals Forces. \emph{J. Colloid Interface Sci.} \textbf{144}, 412 -- 433.
\bibitem[20]{b20}\textsc{Zhao, H. \& Beysens, D.} 1995 From droplet growth to film growth on a heterogeneous surface:
condensation associated with a wettability gradient. \emph{Langmuir} \textbf{11}, 627 -- 634.
\end{thebibliography}
\end{document}